\documentclass{aa}  

\usepackage{graphicx}
\usepackage{txfonts}
\usepackage{pdflscape}
\usepackage{soul}

\usepackage[colorlinks=true,urlcolor=blue,citecolor=blue,linkcolor=blue]{hyperref}

\graphicspath{{./}{figures/}}

\defcitealias{jerabkova_2021}{J21}

\newcommand{\zh}{z_{m,\text{h}}}
\newcommand{\Rmax}{R_\text{max}}

\newcommand{\RSun}{\text{R}_\odot}

\begin{document}

   \title{Impact of the Galactic bar on tidal streams within the Galactic disc}
   \subtitle{The case of the tidal stream of the Hyades}

   \author{Guillaume F. Thomas\inst{1,2}, Benoit Famaey\inst{3}, Giacomo Monari\inst{3}, Chervin F. P. Laporte\inst{4,5,6}, Rodrigo Ibata\inst{3}, Patrick de Laverny\inst{7}, Vanessa Hill\inst{7}, and Christian Boily\inst{3}}

   \institute{Instituto de Astrof\'isica de Canarias, E-38205 La Laguna, Tenerife, Spain \and Universidad de La Laguna, Dpto. Astrof\'isica, E-38206 La Laguna, Tenerife, Spain \and Universit\'e de Strasbourg, CNRS, Observatoire Astronomique de Strasbourg, UMR 7550, F-67000 Strasbourg, France \and Departament de F\'isica Qu\`antica i Astrof\'isica (FQA), Universitat de Barcelona (UB), c. Mart\'i i Franqu\`es, 1, 08028 Barcelona, Spain \and Institut de Ci\`encies del Cosmos (ICCUB), Universitat de Barcelona (UB), c. Mart\'i i Franqu\`es, 1, 08028 Barcelona, Spain \and Institut d’Estudis Espacials de Catalunya (IEEC), c. Gran Capit\`a, 2-4, 08034 Barcelona, Spain \and Universit\'e C\^ote d'Azur, Observatoire de la C\^ote d'Azur, CNRS, Laboratoire Lagrange, Bd de l'Observatoire, CS 34229, 06304 Nice cedex 4, France\\
              \email{gthomas@iac.es}
             }

   \date{Accepted for publication on August, $4^{th}$ 2023}

\authorrunning{Thomas G. et al.}
  \abstract
   {Tidal streams of disrupted clusters are powerful probes of the gravitational potential of the Galaxy and they are routinely detected in the stellar halo of the Milky Way. It was recently shown that tidal streams of open clusters can now also be detected within the Milky Way disc. In this work, we highlight the fact that disc tidal streams also provide a powerful new diagnostic of the non-axisymmetric disc potential and may, in principle, provide a new constraint on the pattern speed of the Galactic bar. In particular, we show how the stream-orbit misalignment for an open cluster on a quasi-circular disk orbit in the solar vicinity varies as a function of the position with respect to the bar resonances. The angular shift rises beyond corotation, reaching values as high as $30^\circ$ close to the outer Lindblad resonance (OLR), then dropping again and reversing its sign beyond the OLR. We applied this mechanism to the recently detected tidal stream of the Hyades open cluster and we note that the detected stream stars would be very similar when taking a potential a priori  with no bar or with a fast pattern speed of 55~${\rm km}\,{\rm s}^{-1}\,{\rm kpc}^{-1}$ (or lower than 30~${\rm km}\,{\rm s}^{-1}\,{\rm kpc}^{-1}$). However, we find that candidate stream stars are different than previously detected ones when adopting a potential with a bar pattern speed of $39~{\rm km}\,{\rm s}^{-1}\,{\rm kpc}^{-1}$, which is consistent with the most recent determinations of the actual Galactic bar pattern speed. Previously detected Hyades candidate members would, on the other hand, favour a  barless galaxy or a fast bar of pattern speed $55~{\rm km}\,{\rm s}^{-1}\,{\rm kpc}^{-1}$. Interestingly, the previously reported asymmetry in star counts within the leading and trailing tails of the Hyades tidal stream persists in all cases. Our study conclusively demonstrates  that the effect of disc non-axisymmetries cannot be neglected when searching for tidal streams of open clusters and that current candidate members of the Hyades stream should not be trusted beyond a distance of 200 pc from the cluster. Moreover, our study allows for ideal targets to be provided for high-resolution spectroscopy follow-ups, which will enable conclusive identifications of the Hyades stream track and provide novel independent constraints on the bar pattern speed in the Milky Way.}
 

   \maketitle

\section{Introduction}

Tidal streams from dissolving globular clusters in the halo of the Galaxy have long been known to be an incredibly powerful probe of the gravitational potential of the Galaxy, its dark matter distribution, and the laws of gravitation itself. In principle, such dynamically cold stellar streams  offer an opportunity to directly probe the acceleration field over the extent of the detected streams because when the progenitors are of low mass and dissolve slowly, the ejected stars are lost and subsequently characterised by a low relative energy. This leads to streams that tend to  closely  (although not perfectly) follow the orbits of their progenitors. While the situation may become more complicated for massive progenitors, the picture for low-mass progenitors can also be complicated by perturbations, either from local disturbances \citep[e.g. giant molecular clouds or putative dark matter sub-halos][]{erkal_2015,erkal_2016,amorisco_2016} or by global ones such as the effects brought on by the infall of the Large Magellanic Cloud \citep{erkal_2019,vasiliev_2021,koposov2023,Lilleengen2023} or of the Sagittarius dwarf galaxy \citep{dillamore_2022} or the presence of the Galactic bar at the center of the Galaxy \citep{hattori_2016,pearson_2017,thomas_2020,dillamore_2023}. 

Recently, detailed analyses of $Gaia$ data have allowed  the realm of the study of tidal streams to be extended to those coming from open clusters inside the Galactic disc. Such tidal streams have been detected up to 1 kpc from the  Hyades cluster \citep{oh_2020,jerabkova_2021}, but also around Praesepe, Coma Berenices, or NGC 752 \citep{boffin_2022,kroupa_2022}. Interesting asymmetries have been found in these streams, which could potentially be attributed to close encounters with massive dark matter sub-halos, although such asymmetries are also reminiscent of asymmetries in some globular cluster streams \citep{thomas_2018} and could challenge Newtonian gravity \citep{kroupa_2022}; also, they could argue in favour of a type of dark matter that breaks the weak equivalence principle \citep{kesden_2006a,naik_2020}.

However, such detections assume a priori that the progenitor orbits within an axisymmetric potential, while the Milky Way disc is actually known to harbor a massive central bar component. Here, we investigate in detail the effects that the bar could have on the orientation of such disc tidal streams and how this can affect the selections of candidate stream stars. We show that these selections can, in principle, serve as powerful new probes for constraining the bar pattern speed.

The pattern speed of the Galactic bar has been historically known as being difficult to estimate, with some contradictory indications coming from different data. For a long time, the prevalent value was estimated to be around $\sim 55~{\rm km}\,{\rm s}^{-1}\,{\rm kpc}^{-1}$, following the work of \citet{dehnen_1999,dehnen_2000a} and \citet{fux_2001}, which showed that the prominent Hercules moving group \citep[e.g.][]{dehnen_1998a,famaey_2005a} in local velocity space could be explained if the Sun is placed just outside the $2:1$ Outer Lindblad Resonance (OLR) of the bar \citep[see also][]{antoja_2014}. This was confirmed by an analysis of the local non-axisymmetric Oort constant \citep{minchev_2007} as well as various other lines of evidence \citep{quillen_2011,fragkoudi_2019}. However, subsequent works on the density of red clump stars in the disc \citep{wegg_2015} and on the gas kinematics \citep{sormani_2015,li_2016}, followed by dynamical modelling of the stellar kinematics in the inner Galaxy \citep{portail_2017}, as well as by an analysis of proper motion data from the VVV survey \citep{sanders_2019,clarke_2019} all point to a pattern speed of $\sim 40~{\rm km}\,{\rm s}^{-1}\,{\rm kpc}^{-1}$. {This value was also favoured by \citet{dillamore_2023} to explain the presence of the ridge structure in the energy-angular momentum space of local stars.} \citet{monari_2019a} then showed that the Galactic model of \citet{portail_2017} could reproduce most of the observed features in local velocity space, including the Hercules moving group with a characteristic dependence on azimuth \citep{monari_2019} if the Sun is placed a bit outside of the co-rotation resonance of the bar. To this day, many different possible pattern speeds are still considered for the Galactic bar \citep{trick_2022} and it is conceivable that this pattern speed actually varies with time \citep{hilmi_2020}. Therefore, gaining access to a new independent probe of the pattern speed of the bar in the form of tidal streams of open clusters in the disc is most desirable. 

In the present paper, we first describe the bar potential that we use for our simulations (Sect.~\ref{sec:pot}) before describing the set-up of our simulation of the stream of the Hyades cluster (Sect.~\ref{sec:sim}). In that section, we show that a `fast' rotating bar with pattern speed $\sim 55~{\rm km}\,{\rm s}^{-1}\,{\rm kpc}^{-1}$ yields very similar results to a potential without bar, whilst a slower pattern speed of $39~{\rm km}\,{\rm s}^{-1}\,{\rm kpc}^{-1}$ presents a very strong stream-orbit misalignment. We then apply those models to observations to re-assess the probability of candidate stream members for different bar pattern speeds (Sect.~\ref{sec:discussion}) and we conclude that high-resolution spectroscopy will be needed to disentangle both types of models. Our conclusions are summarised in Sect.~\ref{sec:conclusion}.

\section{Bar potential} \label{sec:pot}

We start by setting $(R,\theta,z)$ as the Galactocentric cylindrical coordinates. The plane $z=0$ corresponds to the Galactic plane. The angle $\theta$ increases in the clockwise sense, and the line $\theta=0$ passes through the Sun and the Galactic centre. We can define corresponding Cartesian Galactocentric coordinates as $x=-R\cos\theta$, $y=R\sin\theta$, so that the Sun lies at $(x,y)=(-\RSun,0)$, if $\RSun$ is the cylindrical distance of the Sun from the Galactic centre.

The model of the Milky Way used in this paper consists of a time-dependent Galactic potential ($\phi_\text{tot}$) composed of a background axisymmetric potential ($\phi_0$) and of three non-axisymmetric components corresponding to the three first even modes ($\phi_m$) of the bar potential, such that:

\begin{equation}
    \phi_\text{tot} (R,\theta,z,t)=\phi_0(R,z) +\sum_{m=2,4,6} \phi_m (R,\theta,z,t).
\label{eq:pot_tot}
\end{equation}
The contribution to the total potential of each mode $m$ of the bar can be computed from the amplitude of the mode with respect to the axisymmetric potential ($A_{m/0}\equiv A_m/A_0$),  as follows:

\begin{equation}
    \phi_m (R,\theta,z,t)=\phi_0(R,0) \ A_{m/0}(R) \  \frac{\cos{[m\, (\theta - \Omega_\text{b}\,t + \alpha_0)]}}{1+[z/\zh(R)]^2}\ ,
\label{eq:pot_m}
\end{equation}
where $\Omega_\text{b}$ is the pattern speed of the bar (positive in the clockwise sense), while the scale height $\zh$ is defined as:

\begin{equation}
    \zh(R)=0.45 R+\zeta_m,
\label{eq:zh}
\end{equation}
and $\zeta_m$ is a height. The relative amplitude $A_{m/0}$ is defined as:
\begin{equation}
    A_{m/0}(R)=K_m~(R/\Rmax)^{a_m-1}(1-R/\Rmax)^{b_m-1},
\label{eq:AM0}
\end{equation}
where $K_m$ is a scale factor, $\Rmax$ is a scale length that we take to be $\Rmax=12$~kpc. 

The initial phase $\alpha_0$ at $t=0$ is the same for all modes composing the bar potential and chosen such that the inclination of the $m=2$ mode compared to the $x$-axis at the present time $t_{\rm now}$ is $28^\circ$ (i.e. $\alpha_0-\Omega_\text{b}t_{\rm now} = -28^\circ$). 

Therefore, the $\alpha_0$ angle will refer to the `inclination of the bar' at the beginning of the simulation. As it is apparent from this equation, we imposed that the phase of the different modes of the bar do not change with time. It is important to note that in the following, we are working within the Galactic plane and thus we  do not use the $z$-dependence that introduces a vertical dimming of the contribution of each mode of the bar with $z$.
The parameters used for each potential mode $\phi_m$ are summarised in Table~\ref{table:params} and are chosen to roughly resemble the first three even modes of the bar potential by \cite{portail_2017}. 

\begin{table}
\caption{Parameters used in the bar potential}              
\label{table:params}      
\centering                                      
\begin{tabular}{c c c c c}          
\hline\hline                        
$m$ & $K_m$ & $a_m$ & $b_m$ & $\zeta_m$(kpc) \\    
\hline                                   
    2 & 0.25 &  1.80 &   5.08 & 0.05 \\     
    4 & 8.40 & 4.08 &  10.70 & 0.025 \\
    6 & 210.41 & 5.96 &  16.06 & 0.05 \\
\hline                                             
\end{tabular}
\end{table}

\begin{figure}
\centering
  \includegraphics[angle=0,clip,width=8.5cm]{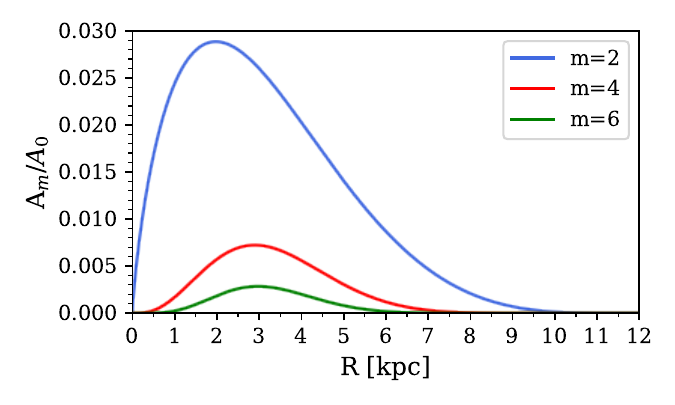}
   \caption{Ratio of amplitude of the different Fourier modes of the bar (m=2, 4, and 6) to $m=0$ along its major axis used in the simulations where the bar is included.} 
\label{fig_amplitude}
\end{figure}

\begin{figure*}
\centering
  \includegraphics[angle=0, viewport= 0 100 580 475,clip,width=17cm]{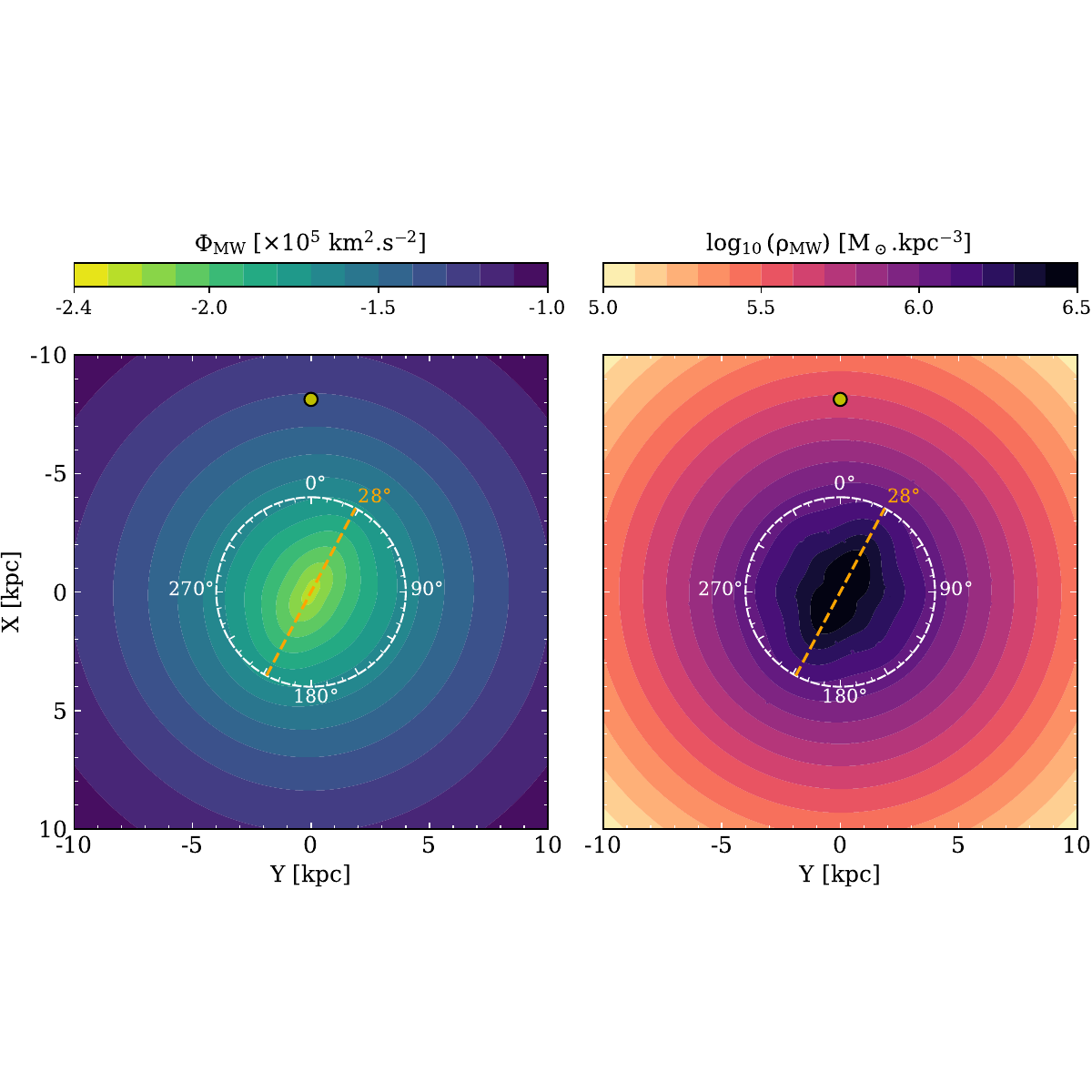}
   \caption{ {\bf Present-day Galactic potential (left panel) and density distribution (right panel) }in the plane of the disc ($z=0$~kpc) of the Milky Way model used by the simulations, with the bar orientated of 28\degr compared to the solar azimuth. In both panels, the representation is Galactocentric, with the current location of the Sun represented by the yellow circle.} 
\label{fig_potential}
\end{figure*}

This model of the Milky Way has been introduced into the {\sc gyrfalcON} N-body integrator \citep{dehnen_2000,dehnen_2002,dehnen_2014} from the {\sc NEMO} stellar toolbox \citep{teuben_1995} as an external acceleration field by modifying the {\sc GalPot} program \citep{dehnen_1998,mcmillan_2017}. In this modified version, which we call {\sc GalBar}\footnote{{\sc GalBar} is publicly available at \url{https://github.com/GFThomas/GalBar.git} }, the classical {\sc GalPot} program is used to compute the background axisymmetric component of the potential. Equation~\ref{eq:pot_m} is used to compute the contribution of each non-zero mode to the total Galactic potential, as expressed in Eq.~\ref{eq:pot_tot}. Then, the corresponding acceleration field is computed locally by measuring the finite difference of potential in a Cartesian grid where the points are spaced in intervals of $10^{-7}$~kpc ($\simeq 20$ AU)\footnote{The choice of this value has been made to avoid propagating rounding errors.}. We verified that this method gives a similar acceleration field than the classical {\sc GalPot} for different axisymmetric potentials.

\section{Simulation of disc tidal streams: the Hyades} \label{sec:sim}

The main objective of this paper is to study the impact of the Galactic bar on the morphology and on the dynamics of stellar streams inhabiting the Galactic disc. One of the best candidates for this analysis is the Hyades tidal stream, due to the proximity of its progenitor with the Sun \citep[45.7 pc,][]{gaiacollaboration_2018a} and for which 6D phase-space information is available for numerous of its stars thanks to the successive $Gaia$ data releases. Using DR2 and eDR3 data, the stream has been measured to be 800 pc long \citep{roser_2019,jerabkova_2021}, a value that is similar to what was predicted in early simulations \citep{chumak_2005,ernst_2011}, in which its progenitor, the Hyades open cluster, had a total mass of 1230 M$_\odot$ at the moment of its formation $600-700$ Myr ago \citep{perryman_1998,lebreton_2001,degennaro_2009,reino_2018,douglas_2019,lodieu_2020}.

\subsection{Simulations setup}
The simulations of the Hyades stream were made with the {\sc gyrfalCON} N-body integrator using the barred potential of the Milky Way implemented in {\sc GalBar,} as described in the previous section. The adopted axisymmetric component is similar to the Galactic potential used by \citet{ibata_2020a} and \citet{thomas_2020} to model the GD-1 and the M92 streams. This potential is composed of the bulge, thin disk, thick disk, and interstellar medium from the first model of \citet{dehnen_1998} and of a dark matter halo following a \citet{navarro_1997} profile, with a virial radius of 206 kpc \citep{cautun_2020}, a concentration of $c = 12$, an oblateness of $q = 0.82$ \citep{malhan_2019}, and a mass of $9.6 \times 10^{11}$ M$_\odot$. With this model, the circular velocity at the solar radius \citep[$\RSun = 8.129$~kpc,][]{gravitycollaboration_2018} is of 229 km~s$^{-1}$, consistent
with the value found by \citet{eilers_2019}. For the simulations where a bar is present, the Galactic potential is computed using Eqs.~\ref{eq:pot_tot} and \ref{eq:pot_m}, with the ratio of amplitude of the different modes with respect to the axisymmetric potential shown on Figure~\ref{fig_amplitude} and with a current bar angle for the bar of 28\degr \ with respect to the solar azimuth \citep{portail_2017}. The face-on view of the potential and of the density of this model at the current time are presented in Fig.~\ref{fig_potential}.

The initial Hyades cluster was modelled with 1230 equal-mass particles, based on the previous work of \citet{jerabkova_2021} (hereafter, J21) and following a Plummer profile \citep{plummer_1911} of a total mass of $M_{\text{cl},0}=1230$ M$_\odot$ and a scale length of $r_\text{s}=2.62$~pc, found using the {\sc mkplum} program included in {\sc Nemo}. We chose to have equal-mass particles, rather than having the individual masses distributed assuming an initial mass function (IMF), as chosen, for example, by \citetalias{jerabkova_2021}. This is driven by our choice of the noncollisional {\sc gyrfalcON} N-body integrator with the goal of studying the general morphology and dynamics of the tidal stream in the presence or absence of a bar, rather than the detailed variations along it, which would require taking into account, for instance, mass segregation in the progenitor \citep{evans_2022}.

For all the simulations, the initial position of the Hyades cluster was computed by integrating backward a point mass from the current position of the cluster over 655 Myr, following the prescription of \citetalias{jerabkova_2021}. The current Galactocentric position and velocity of the cluster are computed from its observed parameters, listed in Table~\ref{tab:param}, assuming that the Sun is slightly above the Galactic plane \citep[z$_\odot=25$ pc,][]{bland-hawthorn_2016} and with the solar peculiar motion adopted from \citet{schonrich_2010}, namely (U$_\odot$, V$_\odot$, W$_\odot$) = (11.1, 12.24,
7.25) km~s$^{-1}$ in the local standard of rest coordinates. However, it is important to note here that the oscillations of the cluster around the Galactic plane were removed in order to have a cluster orbiting in the Galactic plane, such that $z_{\text{cl}}=0$~pc and $v_{z,\text{cl}}=0$~km~s$^{-1}$. The reason behind this choice is that the stream formed on non-planar orbit had a strong vertical dispersion, with non-physical borders of the stream around 40 pc above the Galactic plane. Our investigation tends to suggest that this is due to the scale height of the thin disc used in the simulations, but we did not manage to remove this artificial border completely\footnote{Note: this feature is not caused by our implementation of the acceleration field in {\sc GalBar}, as the acceleration field computed with {\sc GalPot} generates the same effect.}. Therefore, in order to have a realistic stream morphology, we decided to constrain the orbit of the cluster to be in the plane of the Galaxy. This does not significantly impact our study, as \citetalias{jerabkova_2021} noted that the vertical oscillations of the Hyades cluster do not have a significant impact on the morphology of the stream. While neglecting these vertical oscillations leads to a slight overestimation of the response to the bar, we found that this overestimation is actually marginal, namely, at around $\simeq$ 1\% of the amount of torque transfer from the bar to the progenitor, regardless of its pattern speed.

Because the cluster is losing stars with time, its orbit is slightly different from that of a point mass. Therefore, the position of the cluster at the end of the simulation is very slightly different from its present day position. Therefore, in order to have simulations that are directly comparable with the observed Hyades stream, we shifted the final simulation snapshot such that the remnant simulated cluster has the same position as the present-day position of the cluster and we rotated it so that the velocity vector of the simulated cluster was aligned with the observed one. This standard operation ensures that the simulated stream is correctly projected on the sky without changing its internal kinematics.

All simulations were made with the same setup of {\sc gyrfalcON}, with a Plummer softening kernel having a smoothing length of 0.004 pc, a maximum level of refinement of $k_{max}=17$, leading to a maximum time-step  of $2^{-17}=7.6 \times 10^{-7}$~yr and a tolerance parameter of $\theta_0=0.4$, which is smaller than the one usually used \citep[see][]{dehnen_2002}. These parameters were chosen such that an isolated progenitor of the Hyades cluster does not lose more than 20\% of its mass over 700 Myr and to have a relative energy error per time step lower than $10^{-7}$. The choice of parameters used here is also supported by the fact that the simulated cluster has an average mass loss of 0.8 M$_\odot$.yr$^{-1}$ in the non-barred Milky Way potential, similar to what was found by \citetalias{jerabkova_2021} with a fully collisional simulation.

\begin{table}
\centering
\caption{Current dynamical properties of the Hyades cluster from \citet{gaiacollaboration_2018a}.}
\label{tab:param}
\begin{tabular}{lr}
\hline
\hline
Parameter & Value \\
\hline
R.A. & 67.985 deg  \\
Decl. & 17.012 deg \\
Distance & 45.7 pc \\
$\mu_\alpha^*$ & 101.005 mas.yr$^{-1}$ \\
$\mu_\delta$ & -28.490 mas.yr$^{-1}$ \\
V$_{los}$ & 39.96 km~s$^{-1}$ \\
\hline
\end{tabular}
\end{table}

\subsection{Results}
The present-time morphology of the simulated Hyades stream embedded in a barred Milky Way with a wide range of pattern speed and in a non-barred MW is shown in Figure~\ref{fig:moprphology}. 
The extent and the global morphology of the Hyades stream evolving in a barless MW is similar to previous simulations \citep{chumak_2005,ernst_2011,jerabkova_2021}, with a mean inclination of the stream of $\simeq 5\degr$ and the plane being tangent to the Solar azimuth (i.e. the $y$-axis). In this axisymmetric potential, the Hyades cluster is on a near-circular orbit ($e=0.1$), with a pericentre of 7.14 kpc and an apocentre of 8.78 kpc. Because the Hyades cluster is currently not near its apses, the stream is globally aligned with the orbit of the cluster, as is commonly the case for dynamically cold streams formed by the disruption of star clusters (e.g. \citealt{price-whelan_2018,malhan_2018,ibata_2020a}, but see \citealt{thomas_2020} for a counter-example). However, when the Hyades cluster is located between the corotation and OLR of the bar, the stream-orbit misalignment becomes highly significant, reaching deflection angle values as high as 30$^\circ$. If the Hyades were inside the corotation of the bar, that is, in the case of slow-moving bars with $\Omega_\text{b}<29$ km~s$^{-1}$~kpc$^{-1}$, all particles of the stream
would approach the bar at a similar time and receive similar torques \citep{hattori_2016}, leading to a present-day Hyades stream that is globally similar to the case of the MW without a bar. 

For faster bars, it is striking to see that the position, the morphology, the length and the average density of stars along the stream are strongly impacted by the bar pattern speed. Indeed, regarding the length of the stream, it reaches a maximum elongation of $\sim 3.5$~kpc for a pattern speed of $\Omega_\text{b}=35$ km~s$^{-1}$, decreasing down to a few hundred parsecs around $\Omega_\text{b}=50$ km~s$^{-1}$, rising up again for faster bars. This effect, called shepherding by \citet{hattori_2016}, results from the different times at which the
 \onecolumn
\begin{landscape}
\begin{figure*}[!ht]
\centering
  \includegraphics[angle=0, clip,width=25cm]{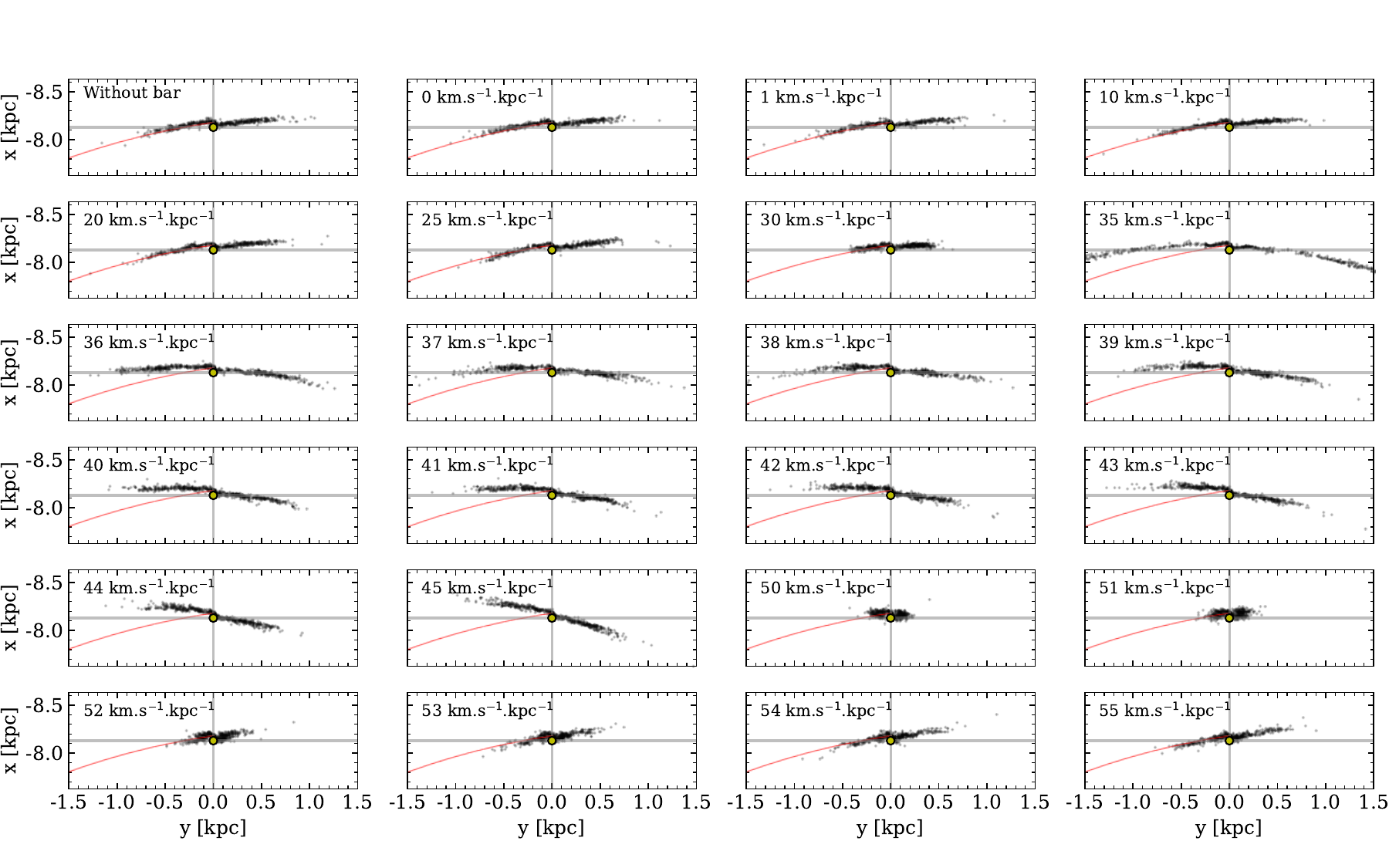}
   \caption{Galactocentric position of the particles of the Hyades-like stream without Galactic bar (upper left corner) and for different pattern speed ranging from $\Omega_\text{b}=$ 0 km~s$^{-1}$ to 60 km~s$^{-1}$. The yellow dots in each panel indicate the position of the Sun and the red line shows the orbit of the cluster.} 
\label{fig:moprphology}
\end{figure*}
\twocolumn
\end{landscape}
\noindent particles along the stream approach the Galactic bar. As a result, the particles along the stream receive different torques from the bar, and the resulting variation of energy is different for each particle. As depicted in \citet{hattori_2016}, the stream’s growth rate depends on the location of the pericentre of the cluster with respect to the bar;  a pericentre near the major axis of the bar will enhance the differences of energy resulting in a growing stream, and on the contrary, while a pericentre near the minor axis of the bar will shrink the stream. 

As already noted, Fig.~\ref{fig:moprphology} clearly shows that the pattern speed of the bar also has an impact on stream-orbit misalignment. The analysis of this specific feature is detailed in the following section.

\subsection{Stream-orbit misalignment} \label{sec:shift}

\begin{figure}
\centering
  \includegraphics[angle=0, clip,viewport= 10 15 420 280,width=8.8cm]{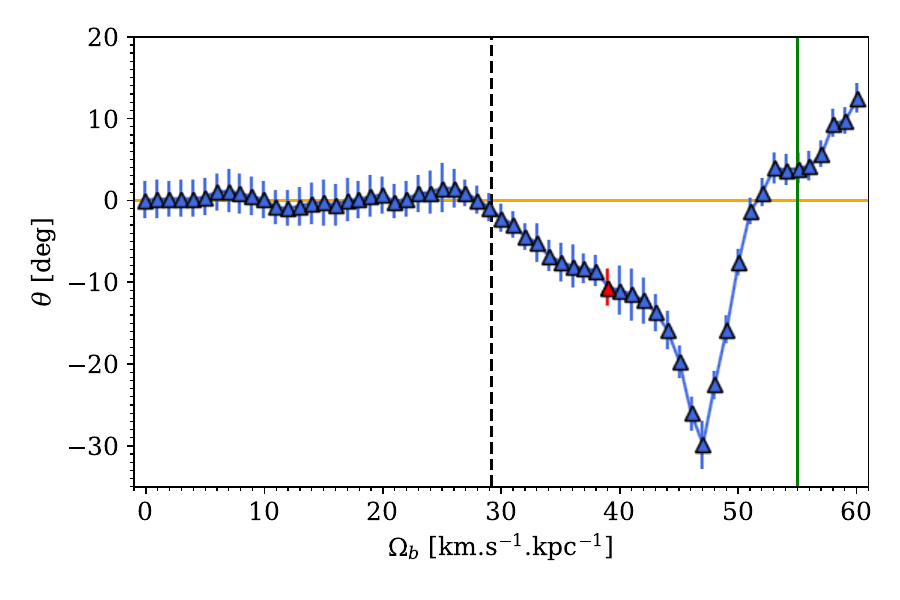}
   \caption{Deflection angle of the plane of the Hyades stream for different pattern speeds of the Galactic bar w.r.t the plane of the stream without a bar. The red triangle highlights the deflection angle for a bar having $\Omega_\text{b}=39$ km~s$^{-1}$~kpc$^{-1}$, and the green vertical line corresponds to $\Omega_\text{b}=55$ km~s$^{-1}$~kpc$^{-1}$. The dashed vertical black line indicates the pattern speed for which the guiding radius of the Hyades cluster will be located at the co-rotation radius ($\Omega_\text{b}\approx 29$ km~s$^{-1}$~kpc$^{-1}$).  The orange line shows the angle in the absence of bar (set to 0\degr \ by definition).} 
\label{fig_angle}
\end{figure}

Figure~\ref{fig_angle} presents the deflection angle ($\theta$), namely, the angle between the track of the stream in presence of a bar with the track of the 
stream embedded in an axisymmetric MW potential, for different pattern speeds between 0 and 60 km~s$^{-1}$~kpc$^{-1}$. The track of the stream was obtained by fitting a straight line on the position of the particles located within a distance of 800 pc from each side of the remnant cluster, and the quoted systematic uncertainties correspond to the difference of inclination when halving the maximum distance to the cluster on each side. A negative deflection angle indicates that the leading arm of the stream is closer to the Galactic centre (i.e.  have higher value along the $x$-axis) than in the case of an axisymmetric MW, and reciprocally for a positive angle. Because the length of the stream varies widely with the pattern speed of the bar -- and because the most extended streams tend to be curved (see the present-day stream for $\Omega_\text{b}=35$ km~s$^{-1}$~kpc$^{-1}$) -- the deflection angle is measured using exclusively the particles within a fixed distance of 800 pc from the cluster.

In this figure, the vertical dashed line corresponds to the pattern speed for which the co-rotation radius is similar to the guiding radius of the centre of mass of the Hyades cluster in the axisymmetric potential ($\Omega_\text{b}\simeq 29$ km \, s$^{-1}$ \, kpc$^{-1}$). Therefore, we can see that when the Hyades cluster is located inside the co-rotation radius (i.e. for $\Omega_\text{b}<29$ km \, s$^{-1}$ \, kpc$^{-1}$), the track of the stream is similar to the case of a stream formed in an axisymmetric MW ($\theta = 0 \degr$).

For a Hyades cluster located outside the co-rotation radius (i.e. $\Omega_\text{b} > 29$ km~s$^{-1}$~kpc$^{-1}$), the deflection angle of the stream track decreases slowly until $\Omega_\text{b}=43$ km~s$^{-1}$~kpc$^{-1}$. For a faster bar, the deflection angle changes drastically, reaching a maximum deflection with an angle of $\theta = -30 \degr$ for $\Omega_\text{b}=47$ km~s$^{-1}$~kpc$^{-1}$. Beyond that value, the deflection angle increases again, reaches $\theta = 0 \degr$ for $\Omega=52 {\rm km} {\rm s}^{-1} {\rm kpc}^{-1}$, up to a plateau of $\theta = 4 \degr$ between $\Omega_\text{b}=53$ and 56 km~s$^{-1}$ \, kpc$^{-1}$, before rising up again for even larger values. This plateau corresponds approximately to the pattern speed of the bar if the guiding radius of the Hyades stream were located at the OLR ($\Omega_\text{b} \approx 50 {\rm km} {\rm s}^{-1} {\rm kpc}^{-1}$).

It is important to note that because stellar streams are spatially coherent over several dynamical timescales \citep{johnston_2008}, a deflection of the position of a stream necessarily implies a deflection of the velocity vectors all along it.

As mentioned in the introduction, several recent works \citep{wegg_2015,sormani_2015,li_2016,portail_2017,bovy_2019,sanders_2019,clarke_2019,monari_2019,tepper-garcia_2021,leung_2023,lucey_2023}, measured the bar pattern speed in the MW to be in the range of $\Omega_\text{b}= 39 - 41$ km~s$^{-1}$~kpc$^{-1}$. For these values, the Hyades stream track will have a non-negligible deflection angle of $\theta \simeq -11 \degr$, which might be in contradiction with the previous detection of candidate stream members, as we discuss in Section~\ref{sec:discussion}. 

\begin{figure*}
\centering
  \includegraphics[angle=0, viewport= 0 40 580 580,clip,width=16.5cm]{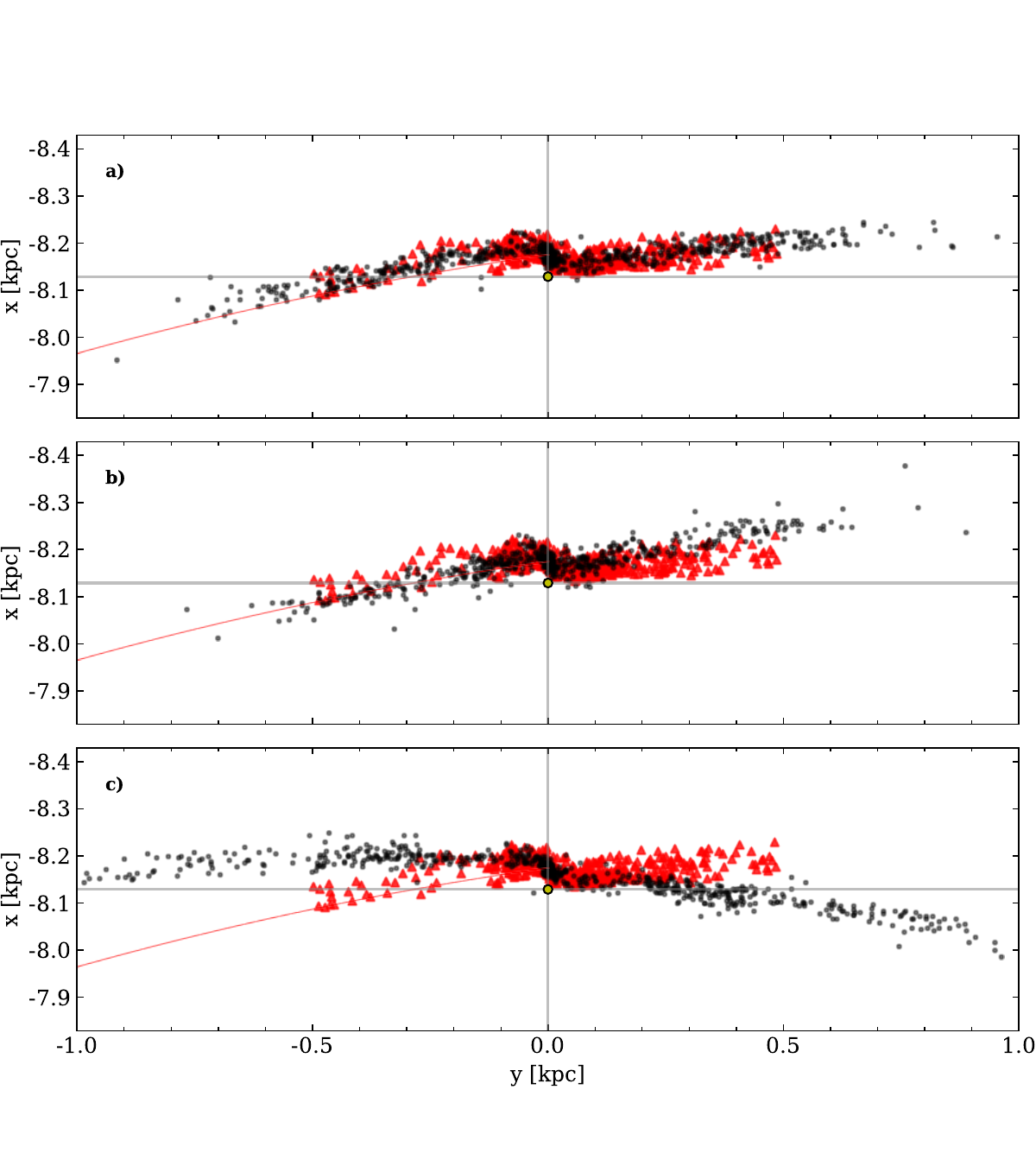}
   \caption{Position of the simulated stream (black points) formed in the case of \textbf{top (a):} a barless Galaxy; \textbf{middle (b):} of a barred MW with a pattern speed of $\Omega_\text{b}=55$ km~s$^{-1}$~kpc$^{-1}$; \textbf{bottom (c):} of a barred MW with a pattern speed of $\Omega_\text{b}=39$ km~s$^{-1}$~kpc$^{-1}$. The red triangles indicate the position of the candidate members of the Hyades stream from the eDR3 sample of \citet{jerabkova_2021}.} 
\label{fig_main}
\end{figure*}

We can see that the Hyades stream has a similar inclination as in the case without a bar for a pattern speed around $\Omega_\text{b}\simeq 52$ km~s$^{-1}$~kpc$^{-1}$. However, for this specific pattern speed, the stream is less extended and thicker than without a bar, as visible on Figure~\ref{fig:moprphology}. A slightly faster bar, with a pattern speed of $\Omega_\text{b} = 55$ km~s$^{-1}$~kpc$^{-1}$ produces a present-day Hyades stream similar in morphology and length as without a bar, although the stream track presents a small deflection angle of $\theta = 4 \degr$. Nevertheless, this deflection is almost three times less than in the case of a stream formed in a $\Omega_\text{b}= 39$ km~s$^{-1}$~kpc$^{-1}$ barred Galaxy. As discussed in the introduction, this result is interesting since this pattern speed value is close to the older measurements \citep[see][for a review]{bland-hawthorn_2016}, and in particular of the measurement of \citet{minchev_2007} of $\Omega_\text{b} = 53 \pm{1.5}$ km~s$^{-1}$~kpc$^{-1}$ using the variation of the Oort $C$ constant. A similar pattern speed has also been estimated, from the position of the Hercules moving group in the local velocity plane \citep[e.g.][]{antoja_2014}. Although (as explained in the introduction) this last measurement is debatable because the Hercules moving group may also be comprised of stars whose orbits are trapped at the co-rotation resonance of a slower bar, which gives the right dependence for its location in velocity space with azimuth \citep{perez-villegas_2017,monari_2019,monari_2019a}.

\begin{figure*}
\centering
  \includegraphics[angle=0, viewport= 8 35 505 190,clip,width=17cm]{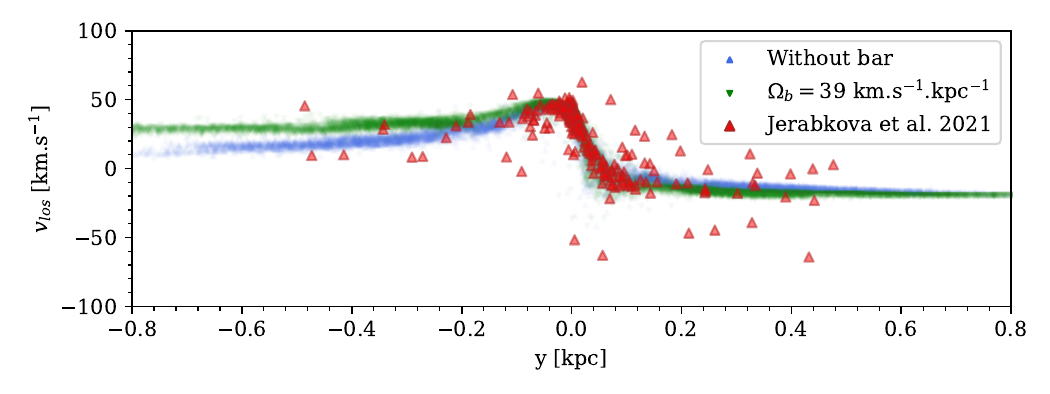}
  \includegraphics[angle=0, viewport= 0 0 505 160,clip,width=17cm]{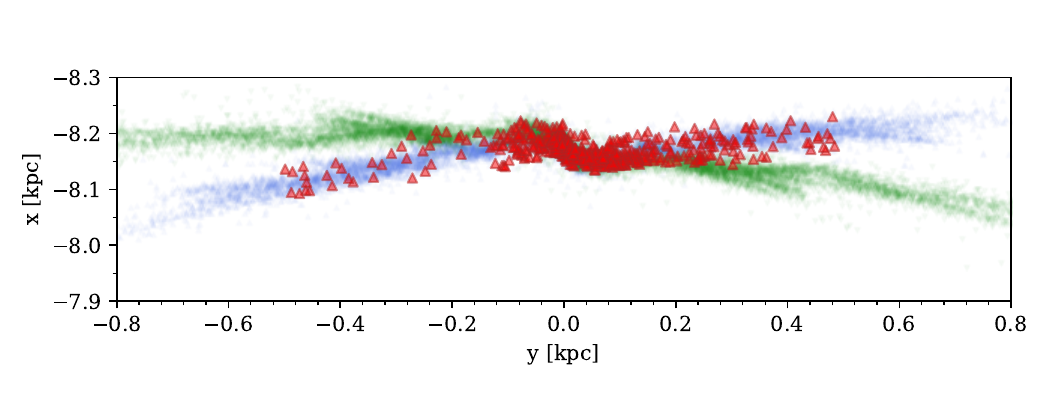}
   \caption{LoS velocity (upper panel) and the Galactocentric position (lower panel) of the Hyades stream simulated in a barless galaxy (in blue) and with a bar having a pattern speed of $\Omega_\text{b}=39$ km~s$^{-1}$~kpc$^{-1}$ (in green). The candidate members of the Hyades stream from the eDR3 sample of \citetalias{jerabkova_2021} are shown by the red triangles.}
\label{fig_obs}
\end{figure*}

\section{Comparison with observations} \label{sec:discussion}

We  undertook a comparison of the simulated streams to the observed Hyades stream and, in particular, to the candidate members selected by \citetalias{jerabkova_2021} using astrometric data from the $Gaia$ early Data Release 3 (eDR3). 

Thanks to the deflection angle of the Hyades stream in presence of a bar, it is in principle possible to use the position and the kinematics of the observed stream to select the best model and, ultimately, to give a constraint on the pattern speed of the Galactic bar. Here, we are particularly interested in three models of the Hyades stream: the streams formed in a MW with a pattern speed of $\Omega_\text{b}= 39$ and $55$ km~s$^{-1}$~kpc$^{-1}$, and the stream formed in a non-barred Galaxy. The reasons behind our choice of focussing on these three models are that for the two barred models, they correspond to the two typical measurements of the pattern speed of the bar in the MW. For the barless model, the justification is that previous state-of-the-art simulations of the Hyades stream were made in an axisymmetric Galaxy model without a bar \citep{chumak_2005,ernst_2011,jerabkova_2021}. For clarity, in the rest of the paper, we refer to the barred-Galaxy with a pattern speed of $\Omega_\text{b}= 39$ km~s$^{-1}$~kpc$^{-1}$ as the 'slow' barred-MW, and to the barred-Galaxy with a pattern speed of $\Omega_\text{b}= 55$ km~s$^{-1}$~kpc$^{-1}$ as the 'fast' barred-MW, which for our present analysis is actually very similar to the 'barless' case.

\begin{figure*}
\centering
  \includegraphics[angle=0,clip,width=10cm]{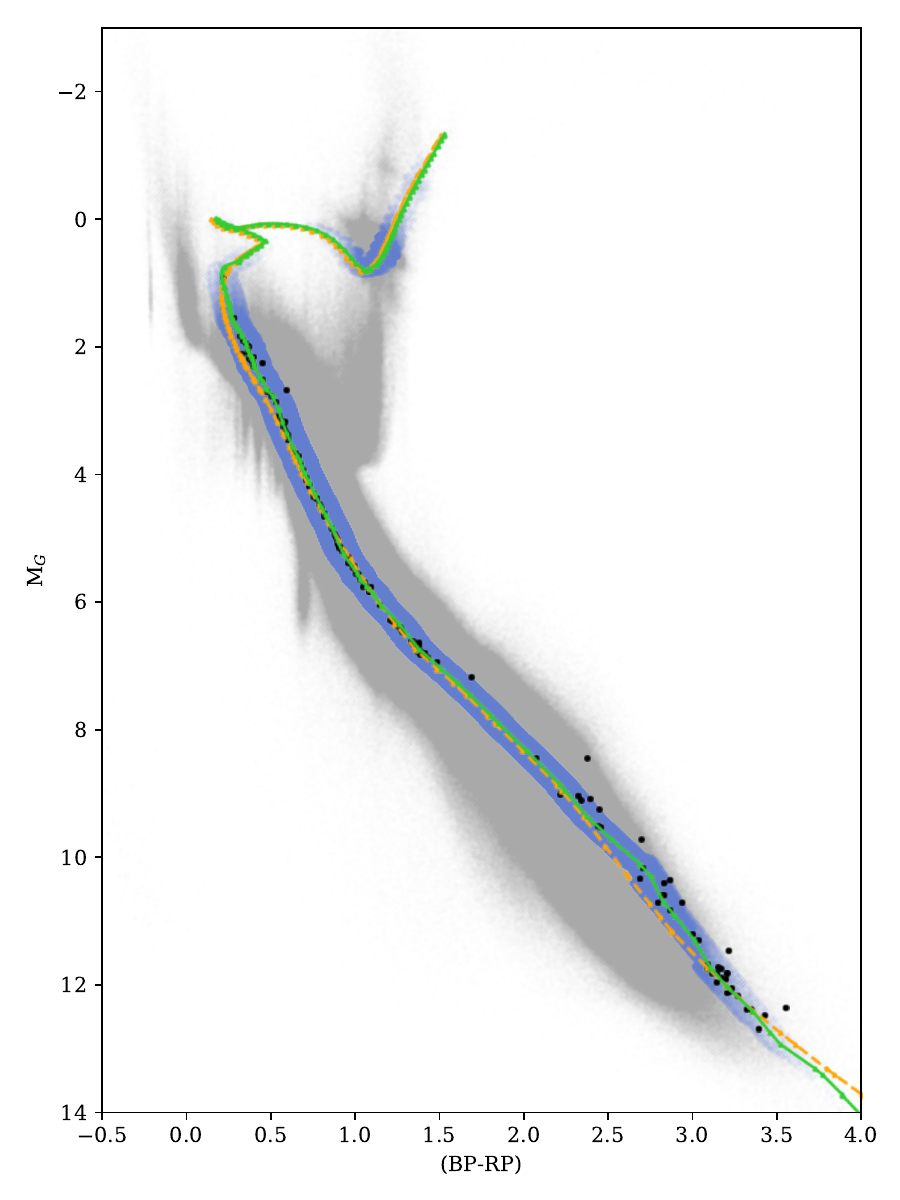}
   \caption{Colour-absolute magnitude diagram of the $Gaia$ DR3 query in grey. On top of it are shown in the orange dashed line and the green plain line, the original and the modified {\sc PARSEC} isochrone of a stellar population of metallicity $Z=0.02$ and age 790~Myr, respectively. The black points correspond to the stars from the \citetalias{jerabkova_2021} selection located in the inner 5 $r_\text{h}$ of the Hyades cluster. The blue shaded region highlights the stars from the $Gaia$ DR3 query that are photometrically pre-filtered. } 
\label{fig_cmd}
\end{figure*}

\subsection{Comparison with the \citet{jerabkova_2021} sample} \label{sec:compare}

Figure~\ref{fig_main} shows the comparison between the Galactocentric position of the simulated streams with the observed eDR3 sample of \citetalias{jerabkova_2021}. As we can clearly see, the candidate members selected in this work are better reproduced by a stream formed in a barless MW, or in a fast barred-Galaxy, than with in a slow barred-MW, especially at a distance beyond $\simeq 150$ pc from the Hyades cluster, where the differences between the models are the strongest. As we already noted in the previous Section, the streams formed in a non-barred Galaxy and in a fast barred-MW are relatively similar. Because the existence of the bar in the Milky Way has been known for decades,  one would at first like to conclude that the observations of the Hyades stream favour a fast rotating bar with $\Omega_\text{b}= 55$ km~s$^{-1}$~kpc$^{-1}$, in contradiction with the latest measurements, but in agreement with the older estimations (see Section~\ref{sec:shift}). However, it has to be noted here that, to make their selection, \citetalias{jerabkova_2021} used an N-body model that was run in an 'axisymmetric' potential. Therefore, it is not completely surprising that the simulations made in the barless MW or in a fast barred-MW better fit these observations, as it might be possible that some actual stars of the Hyades stream are missing and/or that their sample contains a non-negligible number of contaminant stars from the disc. This is especially relevant in the most distant region of the stream, where the differences between the different model are the strongest.

Indeed, the presence of contaminant stars from the disc in the \citetalias{jerabkova_2021} sample is clearly attested by the high dispersion of the line-of-sight (LoS) velocities of those stars, especially in the farthest region of the stream, as shown on the upper panel of Figure~\ref{fig_obs}. The LoS velocities used here were measured by the {\it Radial Velocity Spectrometer} (RVS) instrument onboard of the $Gaia$ satellite, and they were obtained by cross-matching the \citetalias{jerabkova_2021} sample with the $Gaia$ third Data Release \cite[DR3,][]{Gaiadr3} based on the {\sc source\textunderscore id} parameters. Over the 862 stars initially present in the \citetalias{jerabkova_2021} sample, 430 have LoS velocities, most of them contained in the central region of the stream and in the Hyades cluster itself, with a typical individual uncertainty of 2.9 km~s$^{-1}$. Although the majority of the stars beyond $\simeq 150$ pc do not have LoS measurements, the high velocity dispersion of several tens of km~s$^{-1}$ tends to indicate that the \citetalias{jerabkova_2021} sample is dominated by contaminants in that region. 
 
Therefore, due to the method used to select the stars, and due to the presence of a high fraction of contaminants in the furthest parts of the stream, we conclude that the \citetalias{jerabkova_2021} sample is not suited to discriminate which stream models (and, hence, which bar models) are in best agreement with the observations.

 \subsection{Bayesian membership selection}

To circumvent the problem of having to rely on a sample of observed Hyades stars for which the membership selection is biased toward a barless Galaxy, we developed a Bayesian method using the two simulated streams mentioned above to determine the membership probability of each star to belong to the Hyades stream.

The adopted approach is similar to the method described in \citet{thomas_2022}, itself inspired by the methods used to assign a membership probability to the stars of dwarf galaxies \citep{martin_2013a,longeard_2018,pace_2019,mcconnachie_2020,mcconnachie_2020a,battaglia_2022}. These methods assume that the observations are well described by a two-component distribution, describing the distribution of the stars of the Milky Way and of the stars from a given structure (here, the Hyades stream).
Therefore, the likelihood, $p(\mathbf{u}|f_\text{H})$, of a given star with data $\mathbf{u}$ given a fraction of stars present in the Hyades stream $f_\text{H}$ can be defined as
\begin{equation}
p(\mathbf{u}|f_\text{H}) = f_\text{H} p_\text{H}(\mathbf{u}) + (1 - f_\text{H})\ p_\text{MW}(\mathbf{u})\ , 
\label{eq:likelihood}
\end{equation}
where $p_\text{H}$ and $p_\text{MW}$ are, respectively, the probability distributions of the Hyades stream and of the MW foreground and background.

Both $p_\text{H}$ and $p_\text{MW}$ can be decomposed as a spatial projection density ($p_\text{pos}$), a kinematic ($p_\text{kin}$) and a distance ($p_\text{dist}$) components, assuming that each of them are independent of each other. Using the LoS velocities from the $Gaia$ DR3 it would be possible to add an additional component to the likelihood, but we prefer to use them as an independent validation check of our method. Moreover, not all stars have an LoS velocity, so that with this other component of the likelihood, the spatial coverage of the stream will not be uniform and may potentially bias our selection. 

The $Gaia$ data were downloaded from the ESA archive\footnote{\hyperlink{ https://gea.esac.esa.int/archive/}{ https://gea.esac.esa.int/archive/}} using the same {\sc ADQL} query as \citetalias{jerabkova_2021}, except that we relaxed the parallax cut to be $\varpi \geq 1$ mas ({\sc gs.parallax>=1.0}), instead of $\varpi \geq 2$ mas, to be able to detect the stream up to a heliocentric distance of 1 kpc. We also  included the extinction inferred by GSP-Phot Aeneas from the BP and RP spectra \citep[{\sc ag\_gspphot} and {\sc ebpminrp\_gspphot,}][]{delchambre_2023}. This query includes a cut on the renormalised unit weight error ({\sc RUWE}) associated to each source, such as {\sc RUWE< 1.4,} to remove potential non-single sources or sources with problematic astrometric solution. This query results in 38,256,266 objects, among which $\sim 32$ million have a value of the extinction, which we will refer to as the 'initial' sample.  

\subsubsection{Photometric filtering}

In the following section and in the rest of the paper, the $Gaia$ photometry has been corrected from reddening using the extinction values provided on the $Gaia$ ESA archive as mentioned in the previous section. 

Given the Hyades make up a stellar cluster, the colour-absolute magnitude diagram (CaMD) track is aptly described by a {\sc PARSEC} isochrone \citep{bressan_2012} of 790 Myr and of metallicity $Z=0.02$ ([Fe/H]=-0.03), as shown by \citetalias{jerabkova_2021}. Therefore, it is possible to use the CaMD to pre-filter the stars that are potential members of the Hyades cluster and to exclude the obvious contaminants. However, as is visible on Figure~\ref{fig_cmd}, the CaMD of the stars from the \citetalias{jerabkova_2021} sample located in the inner 5 $r_\text{h}$ ($r_\text{h}$=2.62 pc) of the Hyades cluster is not 'perfectly' described by the {\sc PARSEC} isochrone (orange dashed line), especially for magnitudes fainter than  M$_G \sim 10.0$ mag where the isochrone has a steeper CaMD track than the Hyades. For this reason, the isochrone used to filter the initial sample is modified by hand to better fit the CaMD of the Hyades cluster. The modified isochone is shown by the green line on Figure~\ref{fig_cmd}. From the $\sim 38$ million stars on the initial sample, $\sim 19.4$ million of them are located within 0.1 mag of these CaMD tracks and pass the pre-filtering criteria. These stars are highlighted in blue on Figure~\ref{fig_cmd}. We note that without taking extinction into account, the total number of stars selected after this pre-filtering would be similar, but less crowded in the blue (luminous) part of the track and more crowded in the red part. 

It is important to note in this case that other extinction maps are available \citep[i.e.][]{green2019,lallement2019,lallement_2022}, each having different extinction estimates, particularly outside of the Galactic disc \citep[see discussions in][]{andrae_2023}. Therefore, our choice of a specific map here might have a non-negligible impact on the selection made by the photometric pre-filter, in particular for the redder stars (BP-RP$\gtrsim 2$), for which the impact of the extinction is more important. For instance, stars from the selection made by \citet{oh_2020}, with their membership probability estimated $>0.9$ , which reach a maximum distance of 150 pc from the Hyades cluster and a heliocentric distance of up to 180 pc, the extinction of the stars from GSP-Phot Aeneas ranges from $A_G =0$ to 0.9 mag with a typical uncertainty of 0.08 mag, while the same stars have extinction values up to only 0.08 mag using the extinction map from \citet{lallement_2022}, with the difference being the most important for the reddest stars. The extinction values from $Gaia$ GSP-Phot chosen here are (on average) similar to the extinction measured from Planck and to the extinction measured by \citet{schlegel_1998} at a high galactic latitude \citep{delchambre_2023}; it also seems to better catch the region of high extinction compared to other extinction maps \citep[see Section 3.6 of][]{andrae_2023}. We defer the exploration of the detailed effect of the choice of different extinction maps to later studies.

\subsubsection{Distribution of the Milky Way foreground and background}

Because the MW foreground and background distribution largely dominates the signal, it can be empirically determined  from the filtered $Gaia$ data.

The spatial density distribution is made by counting the number of stars in each of the 49,152 {\sc HEALPix} of level 6 \citep{gorski_2005}, which have an average area of 0.92 deg$^2$. This distribution is smoothed with a Gaussian over 10$\degr$ before being normalised.

Along the Hyades stream, the proper motion distribution presents a wide dispersion due to the heliocentric distance gradient of the stream. Therefore, we preferred to use the physical transversal velocities ($v_{\alpha}^*$, $ v_{\delta}$) rather than the proper motions. The distances used to compute the velocities are obtained by inverting the parallaxes. This is possible since our $Gaia$ sample contains only stars with relative precision on the parallax higher than 10 \citep{luri_2018}. Because the kinematic distribution of the Milky Way is spatially dependent, the kinematic distribution is determined locally, at the centre of each {\sc HEALPix}. For each location, the local kinematic distribution is made from the stars located in the 121 nearest {\sc HEALPix}, which at the equatorial coordinate equator typically correspond to all the stars located within a radius of $3.36\degr$. The local kinematic distributions are binned on a fine grid of $0.5$ km~s$^{-1} \times 0.5$ km~s$^{-1}$ bins ranging between -70 and 70 km~s$^{-1}$ in both $v_{\alpha}^*$, $ v_{\delta}$. The size of the bins have been chosen to have a similar size as the velocity dispersion along the stream measured by \citet{oh_2020}, 0.4-0.8 km~s$^{-1}$, which is similar to the dispersion found in our models. This grid is smoothed over 0.5~km~s$^{-1}$ to remove the local kinematic inhomogeneities. This method leads to a smooth, spatially dependent, kinematic distribution of the Milky Way, and is largely inspired by previous works that used a similar method to make spatially dependent colour-(colour)-magnitude diagrams \citep{martin_2013,thomas_2020}.

Similarly, the heliocentric distance distribution of the Milky Way depends on the position on the sky and is built in the same way as the kinematic distribution. The local distribution is made of 100 bins of 10 pc width between 0 and 1 kpc, and smoothed by a 10 pc width Gaussian before being normalised.

\subsubsection{Distribution of the Hyades stream}

For the distribution of the Hyades stream, we explored two scenarios, a Hyades stream formed in a non-barred MW and one formed in a MW with a rotating bar of pattern speed $\Omega_\text{b}= 39$ km~s$^{-1}$~kpc$^{-1}$. For simplicity, in the rest of the paper, we will refer to these two scenarios as the 'barless' and 'slow bar' model, respectively. As the stream formed in a non-barred MW and in a fast-barred MW of pattern speed $\Omega_\text{b}= 55$ km~s$^{-1}$~kpc$^{-1}$ are relatively similar, the conclusions that we will draw for the barless MW will also stand for the fast-barred MW. For each model, the spatial, kinematic and distance distributions of the Hyades stream were built from 20 realisations of the stream formation simulations.

For the spatial density distribution, the 20 simulations are used to estimate the track of the stream on the sky. However, we decided not to use the simulations to constrain the density variation along the stream itself for two reasons: 1) Due to the non-collisional nature of the simulations we carried out, the actual density of stars along the simulated stream might not be realistic. 2) The observations are largely dominated by the MW distribution and the density variation along the stream can be neglected to find the stream members.
Therefore, as for the Milky Way distribution, the sky is split into {\sc HEALPix} at level 6, but the probability of the pixels having at least one simulated particle is set to a constant. To avoid having some gaps due to the discretisation of the simulations that lead to a non-smooth distribution, we smoothed the distribution with a Gaussian that is $0.5\degr$  in width.

The kinematic and distance distributions were obtained in the same manner as for the MW, but using the particles from the simulations.  

 \begin{figure*}
\centering
  \includegraphics[angle=0, clip,width=18.0cm]{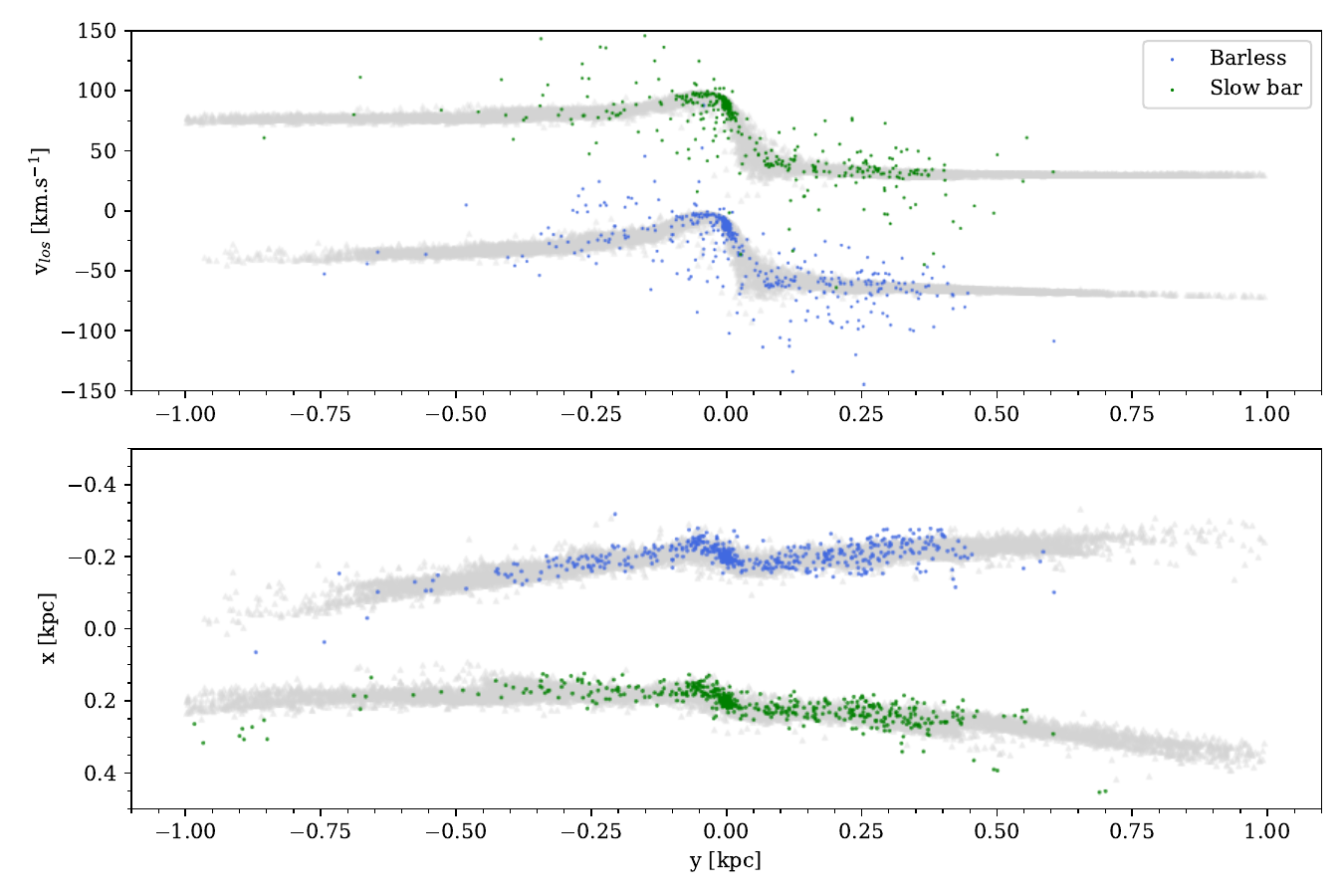}
   \caption{LoS velocity (upper panel) and the Galactocentric position (lower panel) of the candidate member stars selected with our Bayesian method for the barless model (blue points) and for the slow bar model (green points). In both panels, the grey points highlight the track of the simulated stream for each model. The positions of the points have been shifted vertically of $\pm{0.2}$ kpc w.r.t to the current position of the Hyades cluster in the lower panel and of $\pm{50}$ km s$^{-1}$ in the upper panel to separate both selections. } 
\label{fig:pos_p05}
\end{figure*}

\subsubsection{Results}

The fraction of stellar members of the Hyades stream ($f_\text{H}$) was found by exploring the parameter space of the posterior distribution with the {\sc EMCEE} Python package \citep{foreman-mackey_2013}. According to the Bayes theorem, the posterior probability distribution function for the fraction of stars in the Hyades stream $f_\text{H}$, given the filtered data of $N$ stars $D=\{\mathbf{u}_1, ...,\mathbf{u}_N\}$, is $p(f_\text{H}|D) \propto p(D|f_\text{H}) \times P(f_\text{H})$, where $p(D|f_\text{H}) \equiv \Pi_{i=1}^N p(\mathbf{u}_i|f_\text{H})$ is the total likelihood of the data given $f_\text{H}$ and $P(f_\text{H})$ is a flat prior on $f_\text{H}$ between 0 and 1. We found that the posterior distribution of the fraction of stars in the Hyades stream is well approximated near its mode by a Gaussian of mean and standard deviation $(\mu,\sigma)=(82.81,2.72) \times 10^{-5}$ for the barless model, and $(\mu,\sigma)=(165.77,4.89) \times 10^{-5}$ for the slow bar model. From this distribution, it is possible to compute the membership probability of each star of data, $\mathbf{u,}$ of belonging to the Hyades stream, as follows:
\begin{equation}
P_\text{mem}(\mathbf{u}) = \int_0^1  \frac{f_\text{H} \ p_\text{H}(\mathbf{u})}{p(\mathbf{u}|f_\text{H})} \ p(f_\text{H}|D) \ \text{d}f_\text{H}, \ \,
\end{equation}
where $p(\mathbf{u}|f_\text{H})$ is given by Eq.~(\ref{eq:likelihood}).

Applying this method to the photometrically pre-filtered sample led to 580 (569) candidate members with $P_\text{mem} >0.5$ for the barless (slow bar) model, of which 453 (437) have LoS velocity measurements from $Gaia$ {\it RVS}. 327 stars are in common between the two selections, the majority of them being within 150 pc of the cluster. This is not surprising as both models have a very similar distribution in that region, as already mentioned in Section~\ref{sec:compare}. The number of stars for different thresholds of $P_\text{mem}$ for both models are listed in Table~\ref{tab:Nstars}. The barless and slow bar selections have 228(229) stars in common with the sample of $1,003$ stars selected by \citet{oh_2020} that extends up to $\sim 150$ pc too, with their membership probability above 0.9 for almost all of them in both case (99\%). This difference between the number of stars in common with our selection and the total number of stars selected by \citet{oh_2020} is due to the fact that stars of the latter sample not present in ours either do not have Gaia GSP-Phot extinction measurement or are located outside the CaMD selection region. 

The Galactic position and the LoS velocity of these two selections with $P_\text{mem} >0.5$ is shown on Figure~\ref{fig:pos_p05}. A visual inspection of the LoS velocity distribution from the {\it RVS} shows that both selections are far from being devoid of contaminants. 
 We estimated the contamination levels using two methods. For the first method, for each sample, we compared the number of stars confined within a distance of 1~pc along the $y$-axis and of 5 km~s$^{-1}$ in $v_\text{los}$ from a particle in the corresponding set of 20 simulations to the total number of stars with LoS measurement. With this method, we found that the contamination level is of at least 34\% (28\%) for stars located in the stream in the barless (slow-bar) selection. We note that this is a lower limit, since at least one of the two selections must be wrong and should contain more contaminant stars coming from the MW disc, especially far away from the cluster center. In the second method, the contamination fraction is crudely estimated by measuring the fraction of stars of the stream with a velocity compatible at a $3\sigma$ level with the velocity dispersion and the velocity gradient measured by \citet{oh_2020} along the stream. However, this method is valid only up to a distance of 150 pc from the cluster. With this method, we found that the contamination level goes up to 50\% (45\%) for the barless (slow-bar) samples. In comparison, using the same crude method, we found that the \citet{oh_2020} sample would have a contamination fraction of 23\%.
 
 Both selections present a stream with a typical length of 800 pc as found by \citetalias{jerabkova_2021}. The slow-bar selection is more extended than the barless selection, but this can be explained by the fact that the stream, in the simulations on which the former selection is based, is more extended than the latter one. It is particularly interesting to see that  both selections, while different, are still quite well populated, especially in the regions where the two models are clearly distinct (i.e. >150~pc from the cluster). Indeed, in these regions one would expect that the selection based on the model that reproduced the best the morphology and the dynamics of the real Hyades stream would be substantially more populated than the other one; it would be mostly, if not entirely, populated by disc contaminants. However, both samples have the same number of candidates, even in that distant region (237 stars), making it not possible to favour a scenario over the other with only this information at hand.

\begin{figure*}
\centering
  \includegraphics[angle=0, clip,width=18cm]{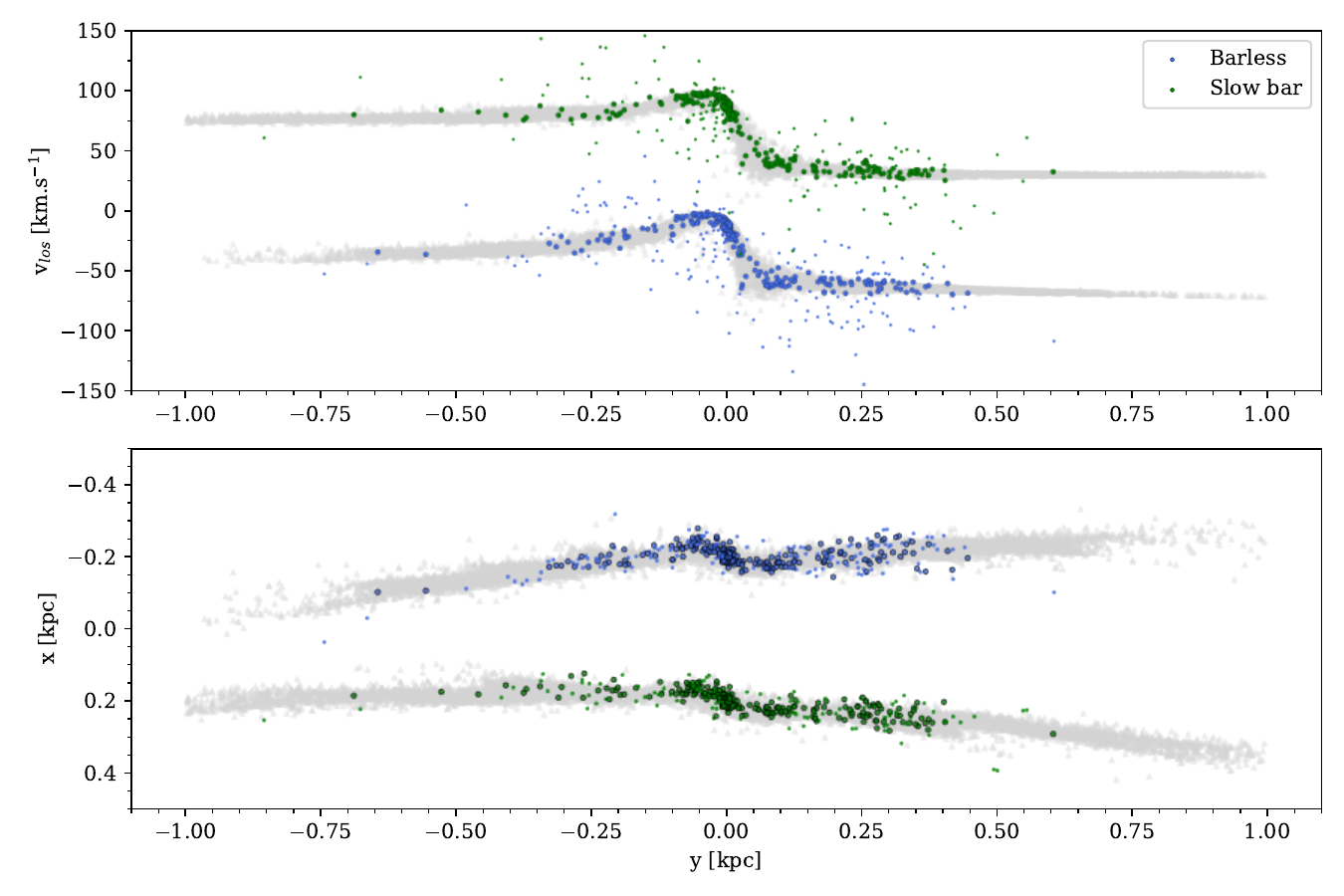}
   \caption{Same as Figure~\ref{fig:pos_p05}, but only for the stars that have a LoS velocity measurement in $Gaia$ {\it RVS} on the lower panel. In both panels, black circles highlight stars that have a LoS velocity compatible with the model on which their selection is based.} 
\label{fig:sel_Vlos}
\end{figure*}

When comparing the morphology and the kinematics of the two selections with the two models they are based on, both possess a significant number of stars that have the expected dynamical characteristics especially at large distances, where the models are the most different. This is particularly striking on Fig.~\ref{fig:sel_Vlos}, where the LoS velocities are used to clean each sample from contaminant stars with the first method described hereabove, that is, by selecting only those stars that are at a maximum distance of 5 km~s$^{-1}$ in $v_\text{los}$ and 1~pc along the $y$-axis of a particle from the corresponding set of 20 simulations. The stars belonging to these two cleaned samples are highlighted by the black circles on Fig.~\ref{fig:sel_Vlos}. Even then, we can see that both cleaned selections still possess a significant number of stars up to $\sim 400$ pc in each arm of the stream, making it impossible to discard or to favour one scenario over the other, especially since both sample have a similar number of stars, even outside the cluster itself.

Our interpretation of the fact that both selections are largely populated by stars that exhibit a kinematic behaviour that is consistent with the two scenarios that we explored is that the number of stars from the disc having similar dynamical properties as the Hyades stream exceeds the number of stars from the stream itself, particularly at large distances from the cluster. Here, it should be additionally noted that the presence of spiral arms, possibly generating the resonant Hyades 'moving group' \citep[e.g.][]{famaey_2007hyades,famaey_2008,pompeia2011,mcmillan2011,mcmillan2013}, could complicate the picture even further. Many disc stars seem to have a similar metallicity to the cluster, as they passed the photometric pre-filter. This explains why in both scenarios, we can detect stars with similar dynamical properties as the Hyades tidal stream, even in regions where the stream simulated within the two scenarios has clearly distinct dynamical signatures. Another caveat here is that the simulations the selections are based on do not capture the full complexity of the actual Hyades stream, due to their non-collisional nature or due to the MW models used here (neglecting the spiral arms, as explained above), which remain rather simple with respect to the true complexity of our Galaxy.

To push our analysis a bit further, we used the calibrated atmospheric parameters derived from the $Gaia$ {\it RVS} spectra by the {\it MatisseGauguin} pipeline to check the consistency with the abundances from the Hyades cluster. Following the guidelines from \citet{recio_blanco2023}, we selected stars having reliable parameters using the {\sc flags\_gspspec}\footnote{See Table~2 from \citet{recio_blanco2023}.} from the $Gaia$ dataset :
\begin{itemize}
\item VbroadT=0
\item VbroadG=0
\item VbroadM=0
\item KMgiantPar=0.
\end{itemize}
The first three parameters assume that the impact of rotation is minimal in the temperature, surface gravity, and metallicity estimates, whilst the last one guarantees that the parametrisation for K and M-giant is correct\footnote{Note: this last parameter is negligible, as our sample should not, in principle, include such stars.} With these criteria, 149(145) stars with a $v_\text{los}$ for the barless (slow bar) selection have a metallicity and $\alpha$-abundance estimate. Among these, 117 stars are included in both selections. A total of 77 (75) of these stars are located within $5 r_h$ ($\sim 13$~pc) of the Hyades center in the barless (slow-bar) sample, of which 74 are in common. These stars have $[\text{M}/\text{H}]=+0.14$ with a dispersion of 0.16 and $[\alpha/\text{Fe}]=-0.01$ with a dispersion of 0.14, fully consistent with the values found in the literature for the Hyades cluster itself, with $[\text{M}/\text{H}]\simeq+0.14$ and $[\alpha/\text{Fe}]\simeq0.0$ \citep{strobel1991,cayrel1997,perryman_1998,pompeia2011}. In the barless (slow bar) selection, only 2 (5) of these stars are further than 300 pc from the cluster, namely, where the spatial and kinematic differences between the two selections are the most important. For the barless sample, 1 out of 2 stars has a metallicity and $\alpha$-abundance similar to the core of the cluster, while this fraction is of 2/5 for the slow bar sample. Obviously, this small number statistics for abundances does not allow us to favour one scenario over the other, but these results nevertheless tend to confirm the hypothesis that many contaminant disc stars with metallicity (and $\alpha$-abundance) values similar to the cluster did indeed pass the photometric pre-filter.

Our analysis shows that to have the ability  to favour a scenario over the other, a high-resolution follow-up of the highly probable member stars of the Hyades stream is required to disentangle the stars originated from the Hyades cluster from the contaminant ones originated from the thin disc. Indeed, the stars of the Hyades cluster present a homogeneous chemical distribution and display individual element abundances that differ from the rest of the thin disc population of similar metallicity \citep{gebran_2010,desilva_2011,pompeia2011,cummings2017}. The most promising element to discriminate which scenario is favoured by the data is lithium: it has been shown by \citet{boesgaard2002} and \citet{pompeia2011} that stars in the Hyades cluster follow a tight Li$-$T$_{eff}$ relation in a narrow temperature range of 5000-6500 K \citep[see Fig.~11 of][]{pompeia2011}, whereas the field stars present a large dispersion in Li-abundance ($\sim 2$ dex) at every temperature. This tight correlation between Li and temperature follows from the fact that stars of the Hyades cluster are coeval \citep{Deliyannis_2000}. We note that this tight correlation is only visible for stars in a narrow range of temperature (5000-6500 K) since outside that range, Li gets destroyed by diffusive or convective downward motions \citep[see][and references therein]{Sestito_2005}. The Li-abundances have also been used to find likely member candidates for 20 other clusters \citep{gutierrez_2020}. Therefore, with a spectroscopic high-resolution follow-up for the stars of our selected samples within a range of $5000 \leq $T$_{eff} \leq 6500$K, it will be possible to check whether the stars selected based on one or the other scenario present a Li$-$T$_{eff}$ relation similar to the one seen in the Hyades cluster, or if it spans a wide range of Li-abundance values that are characteristic of field stars. In particular, it will be interesting to focus this analysis on stars of the trailing arm, located at a distance of 250-300 pc of the cluster, since in that region the candidate members are mutually exclusive due to the different trend in $V_{los}$. Our candidate members in both scenarios will therefore be ideal targets for a WEAVE follow-up study \citep{jin_2022}.

\begin{table}
\centering
\caption{Number of candidate members of the Hyades stream for different threshold of $P_\text{mem}$ for the two models explored. }
\label{tab:Nstars}
\begin{tabular}{|l|c|r|r|}
\hline
Model & $P_\text{mem}$ & $N_\text{stars}$ &  $N_\text{stars}$ with LoS\\
\hline
 & $>$0.5 & 580 & 453 \\
 Barless & $>$0.7 &  415 & 355 \\
 & $>$0.9 &  308 & 260 \\
 \hline
 &$>$0.5 & 569 & 437 \\
Slow bar & $>$0.7 & 420 & 340 \\
 & $>$0.9 & 314 & 259 \\
\hline
\end{tabular}
\end{table}

\subsubsection{Leading and trailing arm asymmetry}

\begin{table}
\centering
\caption{ Number of stars of the Hyades stream located at a range of 50-200 pc from the cluster in the leading ($N_\text{lead}$) and trailing ($N_\text{trail}$) arms, and the derived number ratio for the sampled made from the two scenario we explored and for different criteria, as well as for the \citetalias{jerabkova_2021} sample.}
\label{tab:Q}
\begin{tabular}{|l|l|c|c|c|}
\hline
Dataset & Criteria & $N_\text{lead}$ &   $N_\text{trail}$ & $q_{50-200\text{pc}}$\\
\hline
  & $P_\text{mem}>0.5$ & 86 & 57 & 1.51 $_{\pm{0.25}}$\\
Barless  & $P_\text{mem}>0.8$ & 62 & 42 & 1.48 $_{\pm{0.30}}$\\
  & $V_\text{los} $ cleaned & 76 & 55 & 1.63  $_{\pm{0.40}}$\\
\hline
  & $P_\text{mem}>0.5$ & 71 & 58 & 1.22 $_{\pm{0.22}}$\\
  Slow bar  & $P_\text{mem}>0.8$ & 56 & 41 & 1.37$_\pm{0.28}$\\
  & $V_\text{los} $ cleaned & 42 & 31 & 1.35$_{\pm{0.32}}$\\
\hline
  & All & 158 & 61 & 2.59 $_{\pm{0.39}}$\\
  \citetalias{jerabkova_2021}  & $V_\text{los} $ cleaned & 40 & 14 & 2.86 $_{\pm{0.89}}$\\
\hline
\end{tabular}
\end{table}

Previous observations of the Hyades stream have found that the trailing arm is less extended and less populated than the leading arm \citep{roser_2019,jerabkova_2021}, possibly indicating either a disturbance from a putative massive dark matter sub-halo or a departure from the standard gravitational framework \citep{thomas_2018,kroupa_2022} in accordance with MOND \citep{milgrom_1983,famaey_2012}. For instance, based on the \citetalias{jerabkova_2021} sample, \citet{kroupa_2022} measured that the ratio between the number of stars located in a distance range of 50-200 pc from the cluster in the leading ($N_\text{lead}$) and in the trailing ($N_\text{trail}$) arms is of $q_{50-200\text{pc}}  \equiv N_\text{lead}/N_\text{trail} = 2.53 \pm{0.37}$, while the simulations of this stream made in Newtonian dynamics found a ratio that did not exceed 1 by more than 3\%. 

Next, we aim to revise the ratio of the number of stars in the leading or trailing arms in the selections of Hyades stream candidates that we established previously. As in \citet{kroupa_2022}, we focus on the region in a distance range of 50-200~pc from the cluster in both arms. However, here, we used the distance of the stars along the stream -- and not the projected 2D distance -- to take into account the fact that the Hyades stream that is formed in the barless scenario is slightly wider than that formed in the slow bar scenario within that distance range. Nevertheless, this small modification does not significantly change the results, since we measured a number ratio of $q_{50-200\text{pc}}  = 2.59 \pm{0.39}$ for the \citetalias{jerabkova_2021} sample, similar to the value reported in \citet{kroupa_2022}. 

As we report in Table~\ref{tab:Q}, for both selections, we found a number ratio lower than with the \citetalias{jerabkova_2021} sample, of $q_{50-200\text{pc}}  = 1.51 \pm{0.25}$ and $ 1.25 \pm{0.22}$ for the selection based on the barless-and-fast-bar and on the slow bar scenario, respectively. This difference can be explained by the lower number of stars that our method selected in the leading arm. Even when selecting stars with the highest probability of membership, the number ratio is lower than previously measured. However, this does not change the conclusion drawn by \citet{kroupa_2022} since a value of $q_{50-200\text{pc}} $ ranging between 1.2-1.6, as we found in our selections, is closer to the number ratio they measured in the simulation of a Hyades-like stream made in the MOND framework (see their Figure 13), and is systematically above the typical Newtonian value. Yet it is interesting to note that the slow bar scenario brings the ratio closer to the Newtonian expectation of 1. We also note that the number ratio measured here is not impacted by the slight difference in heliocentric distance between the two arms of the stream, as the ratios are similar for different ranges of magnitude. However, these results have to be taken with care, as the potential high fraction of contaminant stars and the use of different extinction maps could alter the conclusions drawn in this work.

A final interesting point to note here is that the bar does not produce a clear asymmetry in the leading or trailing arm in the case of the Hyades stream, regardless of its pattern speed, since we found a number ratio around unity in the simulations made with a fast or slow bar, contrary to what we measured in real candidate stars based on those models. This is thus different from what had been found in the case of the Palomar 5 stream, whose asymmetry had been tentatively attributed to the bar \citep{pearson_2017}.

\section{Discussions and conclusions}\label{sec:conclusion}

In this work, we applied a new implementation of a multi-modal Galactic bar in the {\sc GyrfalcON} $N$-body code to the simulation of the tidal stream of the Hyades open cluster. We analysed how the Galactic bar affects the morphology and the dynamics of such a stellar stream inhabiting the solar vicinity on a low-eccentricity disk orbit. The (collisionless) $N$-body simulations of the Hyades stream formation shows that its length and (more importantly) its orientation with respect to the Galactic frame are largely impacted by the pattern speed of the Galactic bar, particularly when the cluster is located between the co-rotation radius and the OLR of the bar. We have not considered a change in the length of the bar when varying its pattern speed, but the most important quantities for our results to hold are the amplitudes of the different modes close to the Sun's position  -- and not their shape in the inner Galaxy. Also, our model does not include the effect of spiral arms, which could actually have a non-negligible effect and it does not consider possible rapid changes in the bar's pattern speed. On the other hand, secular changes in the pattern speed related to such effects as dynamical friction with the dark matter halo would happen on timescales that are much longer than the lifetime of typical open clusters such as the Hyades.

From these simulations, we found that the stream is more inclined toward the Galactic centre in the presence of a slow bar having a pattern speed of $\Omega_\text{b} = 39$ km~s$^{-1}$~kpc$^{-1}$, consistent with the latest direct measurements, than in the absence of a bar. Moreover, the simulations also show that a Hyades stream formed in a fast bar model with $\Omega_\text{b} = 55$ km~s$^{-1}$~kpc$^{-1}$ is comparable in length and inclination to one formed in a barless Galaxy.

The comparison of the simulations with the sample of Hyades candidates selected by \citet{jerabkova_2021} shows that these selected stars favour a fast bar scenario. However, using radial velocity measurements from the $Gaia$ {\it RVS} instrument, we found that the sample of \citetalias{jerabkova_2021} presents a high number of possible disc contaminants, especially in the most distant region of the stream, where the difference of position generated by the pattern speed is the strongest. The method used to select these Hyades candidates is model-dependent, which is biased toward a stream having a very small stream-orbit misalignment, that is, as measured in the presence of a fast bar or in a barless Galaxy. 

Therefore, in the second part of the paper, we present a Bayesian method to select member candidates of the Hyades stream using the astrometry and photometry from the $Gaia$ DR3 based on two scenarios, a barless (equivalent to fast-barred) and a slow barred MW. 
Both selections present a stream with a typical length of 800~pc and they are both populated by a similar number of candidates in the regions where the two selections
are clearly distinct (i.e. each has 237 candidates at $d$>150 pc from the cluster). This, in turn, does not allow to favour one scenario over the
other, a problem related to the presence of a non-negligible number of residual disc contaminants in both cases, which is confirmed by the metallicities and alpha-abundances measurements from $Gaia$ {\sc GSP-spec}.

Unfortunately, the LoS velocities that are available for a large number of stars cannot remove all disc contaminants, especially beyond distances $\sim$ 150 pc from the cluster. Because many of these contaminants have similar photometry and kinematics as the stars of the Hyades tidal stream, we argue that a high-resolution follow-up of our candidate members is required to decrease the contamination from the disc population to increase the purity of the Hyades stream sample and to disentangle which stream track is the correct one, as the Hyades cluster has some peculiar individual abundance ratios with respect to the thin disc \citep{gebran_2010,desilva_2011,pompeia2011}. In particular, stars in the Hyades clusters present a tight Li-T$_{eff}$ relation in the range of $5000 \leq $T$_{eff} \leq 6500$K, whereas field stars span a wide range of $\sim 2$ dex at every temperature \citep{pompeia2011}. Our study indicates that measuring this Li-abundance for stars at a distance of 250-300 pc from the cluster within the trailing arm region, with, for instance, a WEAVE follow-up \citep{jin_2022}, will allow for a conclusive determination of which stream track is favoured, thereby providing novel constraints on the Galactic bar at the same time.

Finally, we note that in almost all cases, we found that the observed Hyades stream has a leading arm which is more populated than the trailing arm, as previously reported \citep{roser_2019,jerabkova_2021,kroupa_2022}. However, the number ratio is lower than measured by \citet{kroupa_2022} using the \citetalias{jerabkova_2021} sample, and is closer to the value they found in their simulations of a Hyades-like stream formed in MOND dynamics.  

While the simulations presented here certainly do not capture the full complexity of the actual Hyades tidal stream (for instance by neglecting the effect of spiral arms on the stream track), they nevertheless demonstrate how complicated the dynamics of streams residing within the Galactic disc can be in comparison to those residing in the stellar halo. Thus, there is a  strong argument for chemistry and dynamics going hand in hand to disentangle this complexity. Our study conclusively demonstrates that current candidate members of the Hyades stream should not be trusted beyond 200 pc from the cluster.

\section*{Acknowledgements}
 The authors thank Alejandra Recio-Blanco and Giuseppina Battaglia for useful discussions, as well as the anonymous referee for comments that increased the scientific quality of the publication. GFT acknowledges support from Agencia Estatal de Investigaci\'on del Ministerio de Ciencia en Innovaci\'on (AEI-MICIN) under grant number PID2020-118778GB-I00/10.13039/501100011033, the MICIN under grant number FJC2018-037323-I, and the AEI under grant number CEX2019-000920-S. BF, GM and R.I. acknowledge funding from the European Research Council (ERC) under the European Unions Horizon 2020 research and innovation program (grant agreement No. 834148) and from the Agence Nationale de la Recherche (ANR projects ANR-18-CE31-0006 and ANR-19-CE31-0017). CL acknowledges funding from the European Research Council (ERC) under the European Union’s Horizon 2020 research and innovation programme (grant agreement No. 852839)

This work has made use of data from the European Space Agency (ESA) mission {\it Gaia} (\url{https://www.cosmos.esa.int/gaia}), processed by the {\it Gaia} Data Processing and Analysis Consortium (DPAC, \url{https://www.cosmos.esa.int/web/gaia/dpac/consortium}). Funding for the DPAC has been provided by national institutions, in particular the institutions participating in the {\it Gaia} Multilateral Agreement. This research has made
use of the SIMBAD database and of the VizieR catalogue access tool, operated at CDS, Strasbourg, France.

The N-body simulations, and the selected candidate members of the Hyades stream will be shared via private communication upon a reasonable request. The {\sc GalBar} code is publicly available at \url{https://github.com/GFThomas/GalBar.git}.

\bibliographystyle{aa}
\bibliography{./biblio}

\end{document}